\documentclass[aps,pra,twocolumn,groupedaddress,showpacs,floatfix]{revtex4}
\usepackage{amssymb}
\usepackage{graphicx}
\usepackage{subfigure}	 	
\usepackage[english]{babel}
\usepackage{float}

\begin{document}

\title{Discrete flat-band solitons in the Kagome lattice}

\author{Rodrigo A. Vicencio$^{1}$ and Magnus Johansson$^{2}$}
\affiliation{$^1$Departamento de F\'{\i}sica, MSI-Nucleus on Advanced Optics, and Center for Optics and Photonics (CEFOP), Facultad de Ciencias, Universidad de Chile, Santiago, Chile}
\affiliation{$^2$Department of Physics, Chemistry and Biology, Link\"oping University, SE-581 83 Link\"oping, Sweden}

\begin{abstract} 

We consider a model for a two-dimensional Kagome-lattice with defocusing nonlinearity, and show that families of localized discrete solitons may bifurcate from localized linear modes of the flat band with zero power threshold. Each family of such fundamental nonlinear modes  corresponds to a unique configuration in the 
strong-nonlinearity limit. By choosing well-tuned dynamical perturbations, small-amplitude, strongly localized solutions from different families may be switched into each other, as well as 
moved between different lattice positions. In a window of small power, the lowest-energy state is a symmetry-broken localized state, which may appear 
spontaneously.

\end{abstract}


\maketitle

Discrete nonlinear systems have developed into an important area of research, with many theoretical and numerical predictions tested and proved experimentally~\cite{reps,Lederer,Morsch}. In particular, nonlinear optics \cite{Lederer} and cold atoms \cite{Morsch} provide excellent implementations of old and new theory, mainly coming from solid state physics. Due to the flexibility of experimental techniques for fabrication of periodic and aperiodic structures, simple as well as very complicated lattice topologies can be controllably obtained in different dimensions.

Generally, when a weak nonlinearity is added to some spatially periodic Hamiltonian system, families of nonlinear localized modes (``gap solitons'') 
bifurcate from the linear band edges \cite{reps}. In generic situations, when the band edge is a non-degenerate local minimum or maximum with non-zero group-velocity dispersion, perturbation theory for weak nonlinearity yields a nonlinear Schr\"odinger (NLS) equation for the slowly varying amplitude of solutions close to the band edge (see \cite{DP12} and references therein). In the most common case of an effective  cubic (Kerr) nonlinearity, one may then conclude from the properties of the corresponding NLS solitons that in one dimension (1D), gap solitons bifurcate from the linear band edge with zero power (norm), while in two dimensions (2D) the bifurcating solution appears at a non-zero power, resulting in a generic {\em excitation threshold} for gap solitons in two (and also higher) dimensions~\cite{FKM97}. In both cases, the soliton envelope decays exponentially with a localization length that diverges in the linear limit.

However, there are some particularly interesting classes of lattices where the above mentioned conditions are not fulfilled. Probably the most well-known example is the 2D Kagome lattice, where one of the tight-binding bands (the lowest-energy one with the sign-conventions used below) is completely flat, and in addition it touches the extremum of the second band at one point so that also the latter becomes degenerate (see, e.g.,~\cite{BWB08} and references therein).  It is therefore an open question, which we here aim at resolving, if and how solitons may bifurcate also from such flat-band linear modes. 

The study of Kagome lattices has a long history, in particular as a prototype system  for geometrically frustrated magnetism (see, e.g.,~\cite{Atwood02}). Recently, successful syntheses of artificial Kagome lattices have been reported in several contexts; e.g., nondiffracting Kagome lattice for light beams were obtained in~\cite{BRD11}, photon-based litography was used to fabricate Kagome lattice structures in~\cite{Wang12}, in~\cite {Jo12} a Kagome optical lattice was realized for trapping ultracold atoms, and in~\cite{Nakata12} a metallic Kagome lattice was fabricated and a flat plasmonic band observed.  Thus, the technology to experimentally observe physical phenomena arising from the presence of a flat dispersion band in a nonlinear lattice appears to be within reach. 

Effects of including interactions in Hubbard-type models on Kagome lattices have been discussed for fermionic~\cite{Fulde10} as well as bosonic~\cite{Altman10} systems, and in both cases an effective gap opening appears between the lowest-energy flat band and the second band at a certain filling factor for sufficiently strong interaction. These gap openings were related to a breaking of translational symmetry of the ground states of the non-interacting lattices.

Concerning localized structures in classical nonlinear Kagome lattices with defocusing nonlinearity, a number of such structures were described in~\cite{Law09}. However, 
these authors focused on complex structures such as vortices and their properties in the limit of strong nonlinearity, and  did not at all discuss the fundamental modes and their connections to the linear flat-band modes (in fact, Ref.\ \cite{Law09} does not even mention the existence of a flat linear band). More recent works discuss defect solitons in Kagome optical lattices with saturable nonlinearity \cite{ZWZ10} and localization of light in Kagome nanoribbons \cite{Molina12}; however, both these works considered exclusively the case of focusing nonlinearity with solitons bifurcating from the edge of the upper band, which is non-degenerate and therefore these solitons follow the standard NLS phenomenology in 2D with excitation threshold~\cite{reps,FKM97}. 

The aim of the present Communication is threefold. Firstly, we present an experimentally
realizable 2D lattice where nonlinear localized modes appear, for a standard Kerr 
nonlinearity, {\em without excitation threshold}. Secondly, we show that this system has, in a window of small power, a {\em symmetry-broken ground state}, 
which may appear spontaneously. Thirdly, we show how small-power, strongly 
localized modes may be {\em moved around} and switched by controllable perturbations. Our specific model uses the Discrete Nonlinear Schr\"odinger 
(DNLS) equation, of direct applicability in nonlinear optics \cite{Lederer}, 
but also relevant in many other physical contexts \cite{reps,Morsch}. 

\begin{figure}[htbp]
\centering
\includegraphics[width=0.44\textwidth]{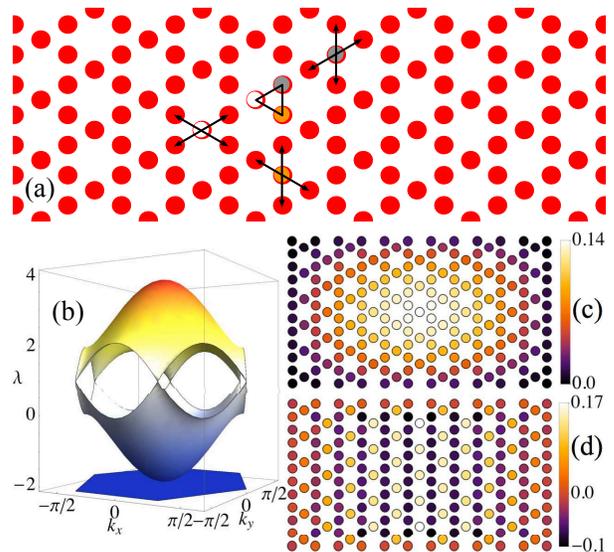}
\caption{(Color Online) (a) Kagome lattice structure with a unit cell of three fundamental sites, including their interactions. (b) Band structure. Linear modes profiles at $\lambda=3.91$ (c), and $\lambda=-1.98$ (d), for a lattice of $205$ sites with rigid boundary conditions.}
\label{f1}
\end{figure}

We thus consider a Kagome lattice with cubic nonlinearity 
using the following equation:
\begin{equation}
i\frac{\partial u_{\vec{n}}}{\partial z}+\sum_{\vec{m}} V_{\vec{n},\vec{m}} u_{\vec{m}}
+\gamma |u_{\vec{n}}|^2 u_{\vec{n}} = 0\ ,
\label{dnls}
\end{equation}
where $z$ corresponds to the normalized dynamical coordinate, 
$\gamma$ to an effective nonlinear cubic parameter, and $u_{\vec{n}}$ represents the field  amplitude at site $\vec{n}$ in a 2D Kagome lattice [see Fig.~\ref{f1}(a)]. The coupling function $\sum_{\vec{m}} V_{\vec{n},\vec{m}} u_{\vec{m}}$ defines the linear interactions between $u_{\vec{n}}$ and its nearest neighbors. Model (\ref{dnls}) possesses two conserved quantities, the norm (power) defined as $P=\sum_{\vec{n}} |u_{\vec{n}}|^2$, and the Hamiltonian (energy) defined as $H=-\sum_{\vec{n}} \left\{\sum_{\vec{m}} V_{\vec{n},\vec{m}} (u_{\vec{m}} u_{\vec{n}}^*+ u_{\vec{m}}^* u_{\vec{n}})+(\gamma/2) |u_{\vec{n}}|^4\right\}$. Unless otherwise stated, the defocusing nonlinear lattice is obtained by fixing $\gamma=-1$, and 
$V_{\vec{n},\vec{m}}=1$ for nearest neighbors and zero otherwise.

Linear solutions ($\gamma=0$) are obtained by solving model (\ref{dnls}) with a stationary ansatz of the form $u_{\vec{n}}(z)=u_{\vec{n}} \exp{(i \lambda z)}$. To obtain the linear spectrum (cf., e.g., \cite{BWB08,Nakata12,Fulde10,Altman10}), we first consider an infinite system and three fundamental sites belonging to a unit cell of the lattice [triangle in Fig.~\ref{f1}(a)]. We construct the corresponding 2D $\vec{k}$-vectors and obtain three different linear bands: 
\begin{equation}
\lambda(k_x,k_y)=-2,\ 1\pm \sqrt{1+8f(k_x,k_y)}\  ,
\end{equation}
where $f(k_x,k_y)\equiv 1+2\cos^{4} (k_x/2)-3\cos^{2} (k_x/2)-\cos^{2} (\sqrt{3}k_y/2) +2\cos^{2} (k_x/2) \cos^{2} (\sqrt{3}k_y/2)$. Fig.~\ref{f1}(b) shows a 3D plot of the band structure in the first Brillouin zone. The upper and lower bands are ``connected'' at $\lambda=1$ by six Dirac points located at the vertices of the hexagon forming the Brillouin zone [$f(k_x,k_y)=-1/8$]. Sketches of the fundamental modes associated to the top of the upper band and to the bottom of the lower band are shown in Figs.~\ref{f1}(c) and (d), for a finite-size lattice with rigid boundary conditions. The largest eigenvalue mode has a typical structure resembling the fundamental mode of any 2D system, where all sites oscillate in phase with a decaying amplitude due to the open boundary conditions. The smallest eigenvalue mode of the lower band possesses a structure reminiscent of
a staggered mode; however, due to the particular geometry of the Kagome lattice it is not possible to construct a fully staggered mode, but a ``frustrated'' state like the one shown in Fig.~\ref{f1}(d). 
Note that, in contrast to the case with periodic boundary conditions, this 
eigenvalue is slightly larger than $-2$ for rigid boundary conditions, and 
therefore this band does not touch the flat band at $-2$ \cite{BWB08}.  
%
%

\begin{figure}[t]
\centering
\includegraphics[width=0.45\textwidth]{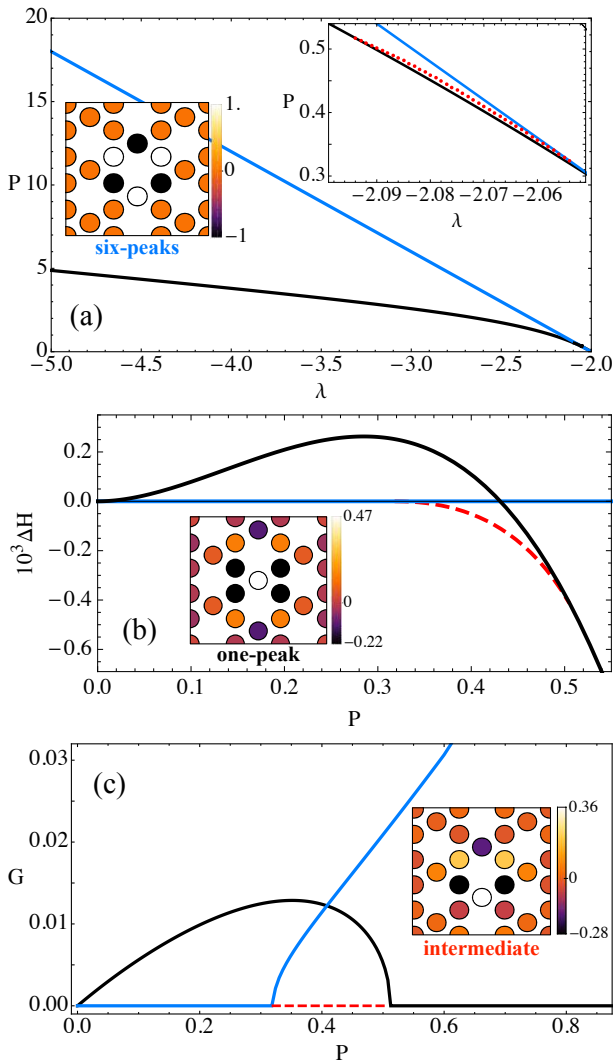}
\caption{(Color Online) (a) $P$ vs $\lambda$, (b) $\Delta H$ vs $P$, and (c) $G$ vs $P$ diagrams. The one-peak, ring, and intermediate solutions are shown with black, blue and red-dashed lines, respectively. Insets show profiles of the stationary modes for $P=0.43$.}
\label{f3}
\end{figure}

The third degenerate-flat band, located at $\lambda=-2$, contains as many states as the number of closed rings in the lattice  (infinite for an infinite system) \cite{BWB08}. These states -- called ``$6$-peaks'' or ``ring'' solutions -- have $6$ peaks with equal amplitude but alternating sign (phase), with strictly zero background [see Fig.~\ref{f3}(a)-inset]. These ring (hexagon) modes constitute ``building blocks'' for a Kagome lattice. Any linear combination of them will generate an exact linear stationary solution of the system. 

Therefore, a fundamental question concerns {\em nonlinear} solutions bifurcating from specific linear combinations of these modes. We first compute this ring stationary solution in the nonlinear regime. It is easily shown, that each ring mode of the flat band can be continued into a nonlinear mode with {\em exactly the same configuration}, only with a frequency shift; thus these solutions are ``exact discrete compactons''~\cite{kecs}. For any nonlinear ring mode, the frequency shift and power are related as $P=6(\lambda+2)/\gamma$. 

For a defocusing nonlinearity ($\gamma<0$), all fundamental nonlinear solutions bifurcate at $\lambda=-2$ ($P=0$) from some linear combination of these ring modes. 
Generally in discrete nonlinear systems, a ``one-peak solution'' is  identified as a family of solutions approaching a single-site localized state in 
the strong-nonlinearity/weak coupling (``anticontinuous'') limit, and 
typically corresponds to the geometrically simplest fundamental nonlinear mode bifurcating from the linear modes at the band edges~\cite{FKM97}. For a Kagome lattice, this mode can be constructed by combining two neighboring rings having one common central site which, in the linear limit, will get twice the amplitude of the other ring sites.
For non-zero nonlinearity, this - compact - solution will no longer be exact; instead, it develops into a discrete soliton with exponentially decaying tails, which continues smoothly to a single-site solution at the anticontinuous limit (larger norms). Therefore, the effective size of this solution will drastically decrease being, for some value of the norm, smaller than the ring solution. We may thus expect an exchange of fundamental properties between the ring and the one-peak solutions if compared at a given norm (cf, e.g., similar features for saturable systems, where multiple changes on the effective size of fundamental solutions result in stability exchanges~\cite{1d,2d,2da}).
%
%

We construct the family of one-peak solutions by implementing a multi-dimensional Newton-Raphson iterative method~\cite{reps}, demanding a (norm-square) accuracy 
of at least $10^{-15}$. Results (including the exact six-peaks solutions) are 
shown in Fig.~\ref{f3}. The $P-\lambda$ diagram 
[Fig.~\ref{f3}(a)] shows how these modes bifurcate from the flat band at $P=0$, i.e., {\em without excitation threshold}.
Moreover, these solutions are very localized for lower values of the norm being, therefore, very unusual 2D discrete solitons originating in the particular topology of this lattice and its fundamental building blocks. [In fact, Fig.~\ref{f3}(a) is more similar to what is obtained for 1D cubic lattices]. For larger norms, these two solutions strongly deviate, the norm content of the ring solution being much larger.  

In Fig.~\ref{f3}(b) we show a $\Delta H-P$ diagram, for $\Delta H\equiv H_i-H_{ring}$ ($i$ represents any solution). For smaller values of the norm, we see how the ring solution ($\Delta H=0$) corresponds to the ground state bifurcating from the linear band at zero norm. Then, we observe a crossing point at $P\approx 0.43$, between the one- and the six-peaks solutions, that indicates an {\em exchange of stability properties}. 

We calculate the linear stability of the -- analytically and numerically -- obtained nonlinear solutions with a standard procedure \cite{sta}: we linearly perturb the nonlinear modes and obtain an equation system for the perturbation. Solving it yields the linear eigenvalue spectrum $\{\omega^2\equiv g\}$ and computing the largest $G = \sqrt{[|g|-g]/2}$ gives the most unstable perturbation mode. A stable (unstable) nonlinear solution corresponds to $G  = 0$ 
($G \neq 0$).
The $G-P$ diagram [Fig.~\ref{f3}(c)] confirms that for small values of the norm, the six-peaks solution is stable (ground state) while the one-peak mode is unstable. For $0.32\lesssim P\lesssim 0.51$ an instability inversion regime with bi-instability appears, so that none of the fundamental solutions correspond to minima in a Hamiltonian representation. Thus, in this 
regime a new stationary solution -- corresponding to a minimum in-between them -- should appear, connecting them in parameter space. For larger power, the one-peak solution is always stable (minimum) while the ring mode is unstable. 

The solution corresponding to a minimum in the bi-unstable region is known as ``Intermediate Solution'' (IS) and constitutes a {\em symmetry-broken stationary solution} appearing when the stability properties of fundamental solutions are exchanged~\cite{pdaub,mika,2d,2da,1d}. In the present case we find a stable IS~\cite{mika,dipolar} in-between two unstable fundamental modes (in other contexts, the opposite is also possible~\cite{pdaub,mika,2d,2da}). In Fig.~\ref{f3} (red-dashed line) we show the appearance of the IS connecting the two fundamental modes. In the stability diagram [Fig.~\ref{f3}(c)] we observe how the IS is stable in the region where the two fundamental modes are unstable simultaneously. Fig.~\ref{f3}(b) shows the emergence of the IS connecting the ring solution with the one-peak mode. In its existence region, the IS possesses the smaller Hamiltonian value and constitutes an effective {\em ground-state} of the system.

\begin{figure}
\centering
\includegraphics[width=0.43\textwidth]{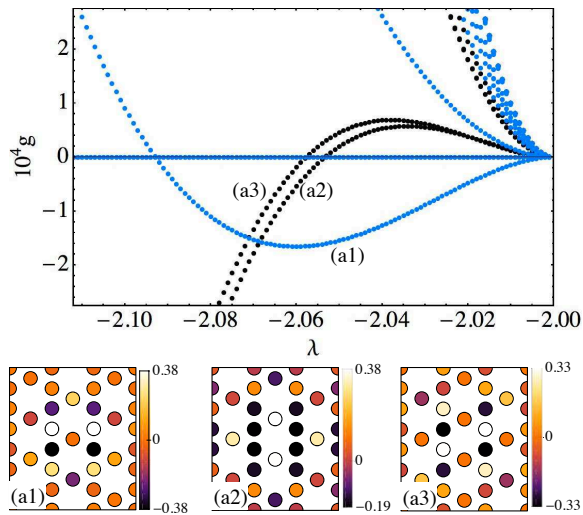}
\caption{Linear spectra $g$ versus $\lambda$ for the one-site (blue dots) and the ring (black dots) solutions in the regime of weak nonlinearity. Insets (a1)-(a3) show the profiles of the indicated linear modes.}
\label{sm}
\end{figure}
%
Fig.~\ref{sm} shows the linear spectra and eigenmodes of the fundamental solutions in the low-power regime. The flat band  spreads out due to the spatial symmetry breaking caused by the particular nonlinear localized mode excited, and isolated localized linear eigenmodes bifurcate from this band. For the one-site solution one such mode is unstable [Fig.~\ref{sm}(a1)], while for the one-ring mode there are two stable (soft) internal modes close to the linear limit [Figs.~\ref{sm}(a2), (a3)]. One of these modes then becomes unstable in the bifurcation where the symmetry-broken intermediate solution is born (the second mode also becomes unstable slightly afterwards). In the linear limit ($\lambda\rightarrow -2$), a gap opening around zero is seen, which is a classical counterpart to the gap openings due to interactions in quantum Hubbard models~\cite{Fulde10,Altman10}.

\begin{figure}[t]
\centering
\includegraphics[width=0.475\textwidth]{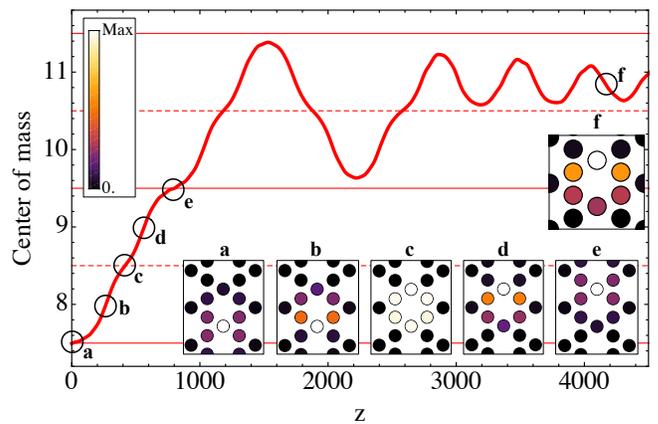}
\caption{(Color Online) Vertical center of mass evolution of an unstable one-peak solution for $P=0.4655$, $k_x=0$, $k_y=0.009$. Insets: zoom of different profiles $|u_{\vec{n}}(z)|^2$. Horizontal full (dashed) lines represent the one-(six-)peaks solutions.}
\label{dina1}
\end{figure}

Exploring the dynamics of unstable one-peak solutions shows how symmetry-broken ground states may appear {\em spontaneously}. The unstable internal mode of the one-peak solution [Fig.~\ref{sm}(a1)] essentially corresponds to a deformation in the vertical direction, similar to a phase gradient in the profile: $u_{\vec{n}} \exp(i \vec{k}\cdot\vec{n})$, with $\vec{k}$ a ``kick'' vector. Fig.~\ref{dina1} shows how applying a small vertical kick yields a slow, smooth movement of a low-norm, very localized solution. The velocity of the center of mass changes
as the effective Hamiltonian (Peierls-Nabarro) potential is traced (cf., e.g., \cite{2d,2da}). The regions with largest velocity correspond to effective potential minima, here corresponding to intermediate stationary solutions with profiles possessing a geometry in-between the one- and six-peaks solutions. 
For the small kick used, the solution jumps coherently one complete site in the vertical direction, and radiation losses prevents it from overcoming the next one-peak barrier (horizontal full lines in Fig.~\ref{dina1}). After oscillations 
through ring modes and intermediate solutions, it 
finally gets trapped with decaying oscillations around one of the symmetry-broken ground states.

%
The possibility of moving very localized solutions across the lattice is certainly important in different physical contexts. Typical 2D nonlinear cubic lattices do not allow mobility of highly localized excitations~\cite{edw2d}, which are thought to be the key entities for controlling the propagation of information -- in the form of waves -- in periodical media. By choosing different phase gradients, we were able to move one- and six-peaks profiles across the whole lattice. For example, giving a vertical kick of $k_y=0.25$ to a one-peak mode of $P=0.502$ resulted in vertical translation of six unit cells (stopping because of borders and radiation effects). While moving, solutions generate some radiation and the dynamics is not as soft as in Fig.~\ref{dina1}. Nevertheless, coherent mobility of highly localized solutions is allowed due to the particular properties of the Kagome lattice. We stress that the flatness
of the linear band, resulting from the Kagome topology, is intimately connected 
to this mobility scenario, since (i) in a generic 2D lattice with dispersive 
bands and Kerr nonlinearity, small-power strongly localized modes will not even 
exist, let alone mobile ones, and (ii) the 
smallness of the Peierls-Nabarro potential results from the small energy shifts 
caused by the nonlinearity-induced lifting of the degeneracy of the flat band. 

In conclusion, we showed, using the Kagome lattice as example, how nonlinear localized modes can bifurcate from a highly degenerate, dispersion-less linear band without excitation threshold. We identified two types of fundamental modes, which were shown to exchange their stability and therefore could be switched into each other through a symmetry-broken intermediate state, constituting the ground state around the exchange region. Since this scenario appears already for a weak (Kerr) nonlinearity, and involves states which are strongly localized due to the flatness of the linear band, it could be of large relevance for practical applications. While for simple cubic regular DNLS lattices symmetry-broken ground states do not appear and stability properties are never exchanged between fundamental modes, the topology of the Kagome lattice allows the appearance of new solutions that can be crucial to improve the dynamical properties of these nonlinear systems, opening new possibilities for controlling the propagation of waves.

The authors thank U. Naether for useful discussions. M.J. thanks the Nonlinear Optics Group, Universidad de Chile, for kind hospitality, and acknowledges support from the Swedish Research Council. This work was supported in part by FONDECYT grant 1110142, Programa ICM P10-030-F, and Programa de Financiamiento Basal de CONICYT (FB0824/2008).

%
%

\end{document}